# CURVE ALIGNMENT BY MOMENTS[1]

By Gareth M. James

*University of Southern California*

A significant problem with most functional data analyses is that of misaligned curves. Without adjustment, even an analysis as simple as estimation of the mean will fail. One common method to synchronize a set of curves involves equating "landmarks" such as peaks or troughs. The landmarks method can work well but will fail if marker events can not be identified or are missing from some curves. An alternative approach, the "continuous monotone registration" method, works by transforming the curves so that they are as close as possible to a target function. This method can also perform well but is highly dependent on identifying an accurate target function. We develop an alignment method based on equating the "moments" of a given set of curves. These moments are intended to capture the locations of important features which may represent local behavior, such as maximums and minimums, or more global characteristics, such as the slope of the curve averaged over time. Our method works by equating the moments of the curves while also shrinking toward a common shape. This allows us to capture the advantages of both the landmark and continuous monotone registration approaches. The method is illustrated on several data sets and a simulation study is performed.

**1. Introduction.** The fundamental paradigm of functional data analysis (FDA) involves treating the entire curve or function as the unit of observation rather than individual measurements from the curve [Ramsay and Silverman (2005)]. As FDA has become more common, many statistical analysis techniques have been adapted to the paradigm. The analysis of functional data possess a number of problems not generally encountered with more classical data. One of the most important is that of misaligned curves. Figure 1 provides a real world illustration of this difficulty using the acceleration curves of ten boys from the Berkeley growth curve study [Tuddenham and Snyder

Received January 2007; revised June 2007.
[1]Supported in part by NSF Grant DMS-07-05312.

(1954)] where the heights of individuals were recorded at regular intervals until age 18. Figure 1(a), which plots smoothed versions of the observed acceleration curves, shows a clear trend of positive and then negative acceleration during the teenage years. However, the onset times, and spread, of these growth spurts can differ by several years so the curves can be considered to be misaligned or "unsynchronized." The dashed line, which represents the cross-sectional mean based on the observed curves, clearly fails to capture the height of the peaks and troughs and underestimates the rate of change in the acceleration curve during the puberty years. Figure 1(b) plots the corresponding curves after synchronization using the approach developed in this paper. Now one can much more clearly discern the general shape of the curves and the gray line, which represents the mean from the synchronized curves, shows that the true peaks and troughs are considerably more extreme than was previously apparent. Computing the mean of a set of curves is only one example of the many applications for which proper alignment of the curves is an essential component. For example, functional principal components analysis [James, Hastie and Sugar (2000) and Rice and Wu (2001)], regression with both functional responses [Zeger and Diggle (1994)] and functional predictors [Ferraty and Vieu (2002) and James and Silverman (2005)], functional linear discriminant analysis [James and Hastie (2001) and Ferraty and Vieu (2003)] and functional clustering [James and Sugar (2003) and Bar-Joseph et al. (2003)] all assume that the starting curves are correctly aligned on the $x$-axis.

The problem of realigning such curves has been studied under different names in several fields. In the statistics literature it is referred to as curve registration [Silverman (1995) and Ramsay and Li (1998)] or, in the context of computing an average curve, structural averaging [Kneip and Gasser (1992)]. It is also called curve alignment in biology and time warping in engineering [Sakoe and Chiba (1978)]. Any set of curves can be decomposed into "amplitude" functions, which measure differences in the $y$-axis, and "warping" functions, which measure differences in location on the $x$-axis. Synchronization requires estimation of the warping functions.

A number of approaches have been proposed for this problem. Marker, or landmark, registration [Kneip and Gasser (1992)] involves selecting common features in the data, such as peaks or troughs, and transforming time so that these features occur together. This method can work well when such features can be easily identified, but tends to perform poorly if no obvious and consistent landmarks exist. In addition, the landmarks often need to be manually identified, preventing the implementation of a fully automatic approach. An alternative method involves aligning curves using a target function. Silverman (1995) proposed registering curves using a simple shift in time such that the average squared distance between each curve and a target function is minimized. This idea was extended in Ramsay and Li (1998) using a Procrustes type fitting procedure on a general nonlinear class of time transformations to provide maximal alignment to the target function subject to suitable smoothness of the transformations. This approach, called "continuous monotone registration," is often very effective but depends heavily on the target function. Generally the cross-sectional mean is used, which can provide misleading results if the curves are significantly misaligned. Other recent work in this area includes Kneip et al. (2000), Rønn (2001) and Gervini and Gasser (2005).

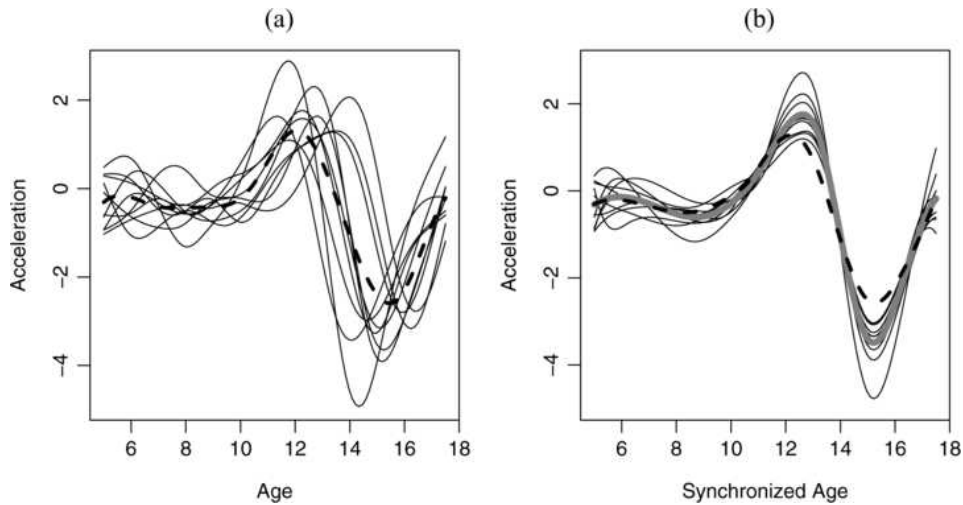

FIG. 1. *Ten acceleration curves from the Berkeley growth curve study,* (a) *unsynchronized curves and* (b) *after alignment. The dashed lines represent the cross-sectional mean based on the observed curves while the gray solid line corresponds to the mean from the synchronized data.*

The aim of this paper is to develop an automatic synchronization method that incorporates the best properties of both the landmark and continuous monotone registration approaches. We start by calculating "moments" for each curve. These moments are intended to capture the locations of important features which may represent local behavior, such as maximums and minimums, or more global characteristics, such as the slope of the curve over time. We then synchronize the curves by both, equating the moments for each curve, which has the effect of aligning common features, and simultaneously shrinking toward a common shape. In situations where obvious marker events are present, our approach has the same advantages as landmark registration. Additionally, when there are no events but an accurate target function can be estimated, we will achieve similar performance to the continuous monotone registration method. However, we show through the use of theory, simulations and real world examples that even in situations where the landmark and continuous monotone registration procedures fail, that is, where obvious markers do not exist and an accurate target function can not be computed, our moments based method can still perform well.

General definitions for the moments of an arbitrary function are developed in Section 2. These moments are defined in terms of "feature" functions which can be designed to detect both local and global characteristics of the curves. In Section 3 we provide a model for the observed or unsynchronized curves. The moments from Section 2 are included as a fundamental part of the model. We also discuss alternative types of warping functions, both linear and nonlinear. A synchronization procedure to fit our model is presented in Section 4. Our procedure attempts to (a) equate the moments among the curves, and hence, align the common "features" in analogy with landmark registration, and (b) shrink the curves toward a common shape in analogy to the continuous monotone registration approach. We provide an algorithm for implementing our method and demonstrate that the estimates are consistent. Our method is illustrated on the Berkeley growth curve data in Section 5. The results from several simulation studies, comparing the performance of our approach with other synchronization methods, are reported in Section 6. Finally, Section 7 discusses the relationship of our approach to other common methods and suggests some further extensions.

**2. Defining the moments of a function.** In this section we develop definitions for the moments of an arbitrary function, $g$, in analogy with the moments of a random variable. The fundamental idea is that, just as one can define the distribution of a random variable through its moments and equate two different distributions by transforming to equate the moments, we can also define the shape of a function through its moments and synchronize two curves by equating their moments. We first introduce the concept of a "feature function," $I_g(t)$, for $g$ and impose the constraints

$$I_g(t) \geq 0 \quad \text{and} \quad \int I_g(t)\,dt = 1,$$

which ensure that $I_g$ is a weighting function. There are various possible choices for $I_g(t)$. Depending on the properties of our data, we may wish to utilize a function that places high weight on the time points corresponding to local features, such as maximums or minimums, or alternatively use a function that places weight according to more global characteristics such as the slope at a given time.

First, we discuss local approaches where most of the weight is concentrated around the time points corresponding to a specific feature in the data. For example, as $r \to \infty$, $I_g^{\max}(t) \propto (g(t) - \min\{g(t)\})^r$ and $I_g^{\min}(t) \propto (\max\{g(t)\} - g(t))^r$ will respectively concentrate their weight on the global maximum and minimum of $g(t)$. We may wish to search for local, as well as global, maximums and minimums. In this case one could utilize

$$I_g^{\text{local}}(t) \propto \begin{cases} \exp\left(-r\dfrac{|g^{(1)}(t)|}{\sqrt{|g^{(2)}(t)|}}\right), & g^{(2)}(t) \neq 0, \\ 0, & g^{(2)}(t) = 0. \end{cases}$$

This function places maximum weight on points where the first derivative is zero. However, $I_g^{\text{local}}(t)$ is also high for points with a low first derivative but a high second derivative. Thus, the function effectively searches for local maximums or minimums where $g$ is changing most rapidly. As $r \to \infty$, $I_g^{\text{local}}(t)$ will place all its weight on the regions around local turning points.

Finally, we examine $I_g^{(m)}(t)$, which places weights according to the absolute $m$th derivative of the curve, $g^{(m)}$, that is, $I_g^{(m)}(t) \propto |g^{(m)}(t)|$. With $m = 0$, this function puts highest weight on large absolute values of $g$. With $m = 1$, most weight is placed on time points where $g$ has a large slope and would be used when we are most interested in regions where a curve is changing rapidly. Setting $m = 2$ searches for points with greatest curvature etc. $I_g^{(m)}(t)$ can be considered to be searching for global characteristics of a curve because it is likely to spread its mass over all time points.

Then, for a given choice of $I_g$, we define the first moment of $g$ by

$$\mu_g^{(1)} = \int t I_g(t)\, dt$$

and the $k$th central moment by

$$\mu_g^{(k)} = \int (t - \mu_g^{(1)})^k I_g(t)\, dt, \qquad k \geq 2.$$

$\mu_g^{(1)}$ provides a measure of the center of $g$ on the time axis, while $\mu_g^{(2)}$ measures variability in $g$. Note that the variability is measured in relation to the time axis and not the $y$, or amplitude, axis. A curve could vary significantly in the $y$-axis, but still have a low value for $\mu_g^{(2)}$. In general, $\mu_g^{(1)}$ will be more useful than the higher order moments when using feature functions such as $I_g^{\max}$ or $I_g^{\min}$ that concentrate on local features. The higher-order moments, that is, $\mu_g^{(k)}$ for $k \geq 2$, increase in importance when using more global feature functions such as $I_g^{(m)}$.

To better understand the properties of $\mu^{(k)}$, we examine the relationship between the moments of a function $h(s)$ and those of the shape invariant function $h(\frac{s-a}{b})$. In this formulation, $h(s)$ is stretched, about $s = 0$, by a factor $b$ and shifted to the right by $a$. Hence, since $\mu^{(1)}$ is a measure of the center of a function and $\mu^{(k)}$ is a measure of variability about the center, stretching by a factor $b$ should multiply the first moment by $b$ and the $k$th moment by $b^k$. For example, one would expect that $\mu_{h((s-a)/b)}^{(2)}$, which measures the variability of the transformed function, would equal $b^2\mu_{h(s)}^{(k)}$. Similarly, a shift of $a$ should add $a$ to the first moment and leave the higher-order moments, which are centered around the first moment, unchanged. We express this mathematically as

$$\mu_{h((s-a)/b)}^{(1)} = b\mu_{h(s)}^{(1)} + a \quad \text{and} \quad \mu_{h((s-a)/b)}^{(k)} = b^k\mu_{h(s)}^{(k)}, \qquad k \geq 2. \tag{1}$$

Theorem 1 shows that, provided we utilize a certain family of feature functions, these properties will hold.

THEOREM 1. *Suppose that $I_g(t)$ is chosen such that*

$$I_{g((s-a)/b)}(t) \propto I_{g(s)}\left(\frac{t-a}{b}\right), \qquad -\infty < t < \infty, \tag{2}$$

*for all $a, g$ and $b > 0$. Then (1) will hold for any function $h(s)$.*

Condition (2) holds for many large classes of feature functions. In particular, the previously mentioned feature functions all satisfy (2) and, hence, their corresponding moments all possess the desirable properties given by (1).

COROLLARY 1. *When utilizing $I_g^{(m)}, I_g^{\max}, I_g^{\min}$ or $I_g^{\mathrm{local}}$, condition (2) is satisfied, and hence, (1) holds. In addition, (2) is satisfied for any $I_g^\phi(t) \propto \phi(g(t))$ where $\phi(t)$ is an arbitrary function.*

The feature functions we have utilized represent only a few of the possible choices one could utilize. In fact, one of the strengths of our approach is the ability to design functions which best suit one's particular data.

**3. The synchronization model.** Let $Y_1(t), Y_2(t), \ldots, Y_N(t)$ represent the unsynchronized functions or curves with $Y_i$ observed at $t_1, \ldots, t_n$ where $t_j \in [0, T]$. Suppose we select $L$ feature functions, $I_g^1, \ldots, I_g^L$, and associated moments, $\mu_g^{(1,k)}, \mu_g^{(2,k)}, \ldots, \mu_g^{(L,k)}$. Then our synchronization model is given by

$$Y_i(t_j) = Z_i(W_i(t_j)) + \varepsilon_{ij}, \qquad i = 1, \ldots, N, \tag{3}$$

$$\mu_{Z_1}^{(l,k)} = \mu_{Z_2}^{(l,k)} = \cdots = \mu_{Z_N}^{(l,k)} = \mu_{\bar{Y}}^{(l,k)}, \tag{4}$$
$$l = 1, \ldots, L \text{ and } k = 1, \ldots, K_l,$$

where $\mu_{\bar{Y}}^{(l,k)} = \frac{1}{N}\sum_i \mu_{Y_i}^{(l,k)}$ and $Z_i(t)$ represents an "amplitude function," which is stretched on the time axis according to a strictly increasing "warping function," $W_i(t)$. In addition, $\varepsilon_{ij}$ represents i.i.d. random measurement errors with $E\varepsilon_{ij}=0$ and $\mathrm{Var}(\varepsilon_{ij})=\sigma^2<\infty$. Note that we have assumed that the curves are all observed at a common set of points simply for notational convenience. There is nothing in our approach that will prevent it working on curves observed at differing time points.

As with all curve registration methods, (3) has an identifiability problem between $Z_i$ and $W_i$. Landmark registration achieves identifiable results by assuming certain markers align for every curve. We generalize this approach using the moments condition given by (4) which forces the $Z_i$'s to have a common "shape." For example, if $I_g^{\max}(t)$, which searches for global maximums, is chosen as the feature function, then (4) states that the $Z_i$'s have a common shape in as much as their global peaks occur at the same time point and that point is equal to the average of the peaks in the observed curves, $Y_i$. As more feature functions are chosen, (4) forces more alignment in the $Z_i$'s. Landmark registration can be seen as a special case of (4) because $\mu_{Z_i}^{(l,k)}$ can be used to identify specific marker events in each curve, such as peaks or troughs, in which case (4) simply forces an alignment of landmarks. However, $\mu_{Z_i}^{(l,k)}$ can also be used to measure more general and more global curve characteristics such as the $m$th derivative as discussed in Section 2. Note that by equating the moments for each curve to $\mu_{\bar{Y}}^{(l,k)}$ we are assuming that positive and negative warping cancels out, in terms of the moments, when averaged over all curves. Without this assumption, $Z_i$ and $W_i$ will not be identifiable.

We model $Z_i$ and $W_i$ using finite-dimensional basis functions. The amplitude function is modeled as $Z_i(t) = \mathbf{z}(t)^T \boldsymbol{\theta}_i$, where $\mathbf{z}(t)$ is a $p$-dimensional basis function and $\boldsymbol{\theta}_i$ represents the corresponding basis coefficients. In the case of the warping functions, since they are restricted to be increasing, we can, without loss of generality, reparameterize them using

$$(5) \qquad W_i(t) = \gamma_{i0} + \int_0^t \exp(f_i(s)) \, ds,$$

where $\gamma_{i0}$ and $f_i$ are unconstrained. As with the amplitude functions, we model $f$ using a finite-dimensional basis, $f_i(s) = \mathbf{w}(s)^T \boldsymbol{\gamma}_i$, where $\mathbf{w}(s)$ is a $q$-dimensional basis and $\boldsymbol{\gamma}_i$ the corresponding coefficients. Several special cases of (5) can be achieved by appropriately restricting the $\boldsymbol{\gamma}_i$ coefficients. We shall explore two in this paper. The first is the linear warping function $W_i(t) = \alpha_i + \beta_i t$ which is achieved by setting $f_i$ equal to a constant. The second is

$$(6) \qquad W_i(t) = \frac{T \int_0^t \exp(f_i(s)) \, ds}{\int_0^T \exp(f_i(s)) \, ds}.$$

Equation (6) has the often desirable property that $W_i(0) = 0$ and $W_i(T) = T$, which means that time is taken to run over a consistent time period for all curves. We utilize b-spline bases for both $\mathbf{z}$ and $\mathbf{w}$ but, in principle, any finite-dimensional basis will suffice.

**4. Curve alignment.** In this section we detail our curve alignment approach for fitting the model from Section 3.

4.1. *A moments based alignment approach.* The aim in fitting our model is to produce estimated curves, $\hat{Y}_{ij} = \mathbf{z}(W_i(t_j))^T \boldsymbol{\theta}_i$, that accurately approximate the observed curves, $Y_{ij} = Y_i(t_j)$, subject to two constraints. First, the shape of the synchronized curves, $Z_i(t)$, should be as close as possible to that of the original curves. Notice that if $W_i'(t) = 1$ for all values of $t$, then $Z_i(t)$ will have an identical shape to $Y_i(t)$. Therefore, we measure the change in shape by examining the departure of $W_i'(t)$ from 1 using $P(W_i) = (\int \{[W_i'(t)]^{-1} - 1\} \, dt)^2$ and, hence, choose a fit such that $P(W_i)$ is small. We penalize the inverse of $W_i'(t)$ to ensure slopes close to zero, which would imply an extremely high level of warping, are strongly discouraged. Second, the shapes of the $Z_i(t)$'s should be as similar as possible to each other. Differences in the shapes can be measured either by examining variability in the $\boldsymbol{\theta}_i$'s from a target $\boldsymbol{\mu_\theta}$, $P(\boldsymbol{\theta}_i) = \|\boldsymbol{\theta}_i - \boldsymbol{\mu_\theta}\|^2$, or by concentrating on the spread of the moments, $P(\mu_{Z_i}) = \sum_l \sum_{k=1}^{K_l} (\mu_{Z_i}^{(l,k)} - \mu_{\bar{Y}}^{(l,k)})^2$. Hence, we find the $\boldsymbol{\theta}_i$'s, $\boldsymbol{\gamma}_i$'s and the $\boldsymbol{\mu_\theta}$ that minimize

$$(7) \quad Q = \frac{1}{N} \sum_{i=1}^{N} \{\|\mathbf{Y}_i - \hat{\mathbf{Y}}_i\|^2 + \lambda_{\text{sync}} P(\boldsymbol{\theta}_i) + \lambda_{\text{mom}} P(\mu_{Z_i}) + \lambda_{\text{W}} P(W_i)\},$$

where $\lambda_{\text{sync}}$, $\lambda_{\text{mom}}$ and $\lambda_{\text{W}}$ are tuning parameters that determine the impact of each term on the fit. $\lambda_{\text{mom}}$ and $\lambda_{\text{sync}}$ control the balance between the continuous monotone registration and landmark registration methods. Conceptually, setting $\lambda_{\text{mom}} = 0$ and minimizing $Q$ is very similar to the continuous monotone registration method of Ramsay and Li (1998). Alternatively, setting $\lambda_{\text{sync}} = 0$ and minimizing $Q$ provides a type of generalized landmark registration. Note that including $\|\mathbf{Y}_i - \hat{\mathbf{Y}}_i\|^2$ and $P(\mu_{Z_i})$ ensures that both (3) and (4) from our synchronization model will hold.

For fixed $\boldsymbol{\mu_\theta}$, minimizing (7) is relatively simple because we only need minimize $Q$ individually over $\boldsymbol{\gamma}_i$ and $\boldsymbol{\theta}_i$. This suggests the following iterative algorithm:

1. For fixed $\boldsymbol{\mu_\theta}$, minimize $Q$ over $\boldsymbol{\gamma}_i$ and $\boldsymbol{\theta}_i$ for $i = 1$. Repeat for $i = 2, \ldots, N$.
2. Set $\boldsymbol{\mu_\theta} = \frac{1}{N} \sum_{i=1}^{N} \boldsymbol{\theta}_i$.
3. Repeat 1 and 2 until convergence.

Step 1 involves a nonlinear optimization, but can be achieved with relative ease because we only need optimize over each curve individually and the derivatives of $Q$ can be computed analytically. Note that we optimize over the $\mu_{Z_i}^{(l,k)}$ as part of step 1, that is, we do not fix $\mu_{Z_i}^{(l,k)}$ at the previous value of $\boldsymbol{\theta}_i$.

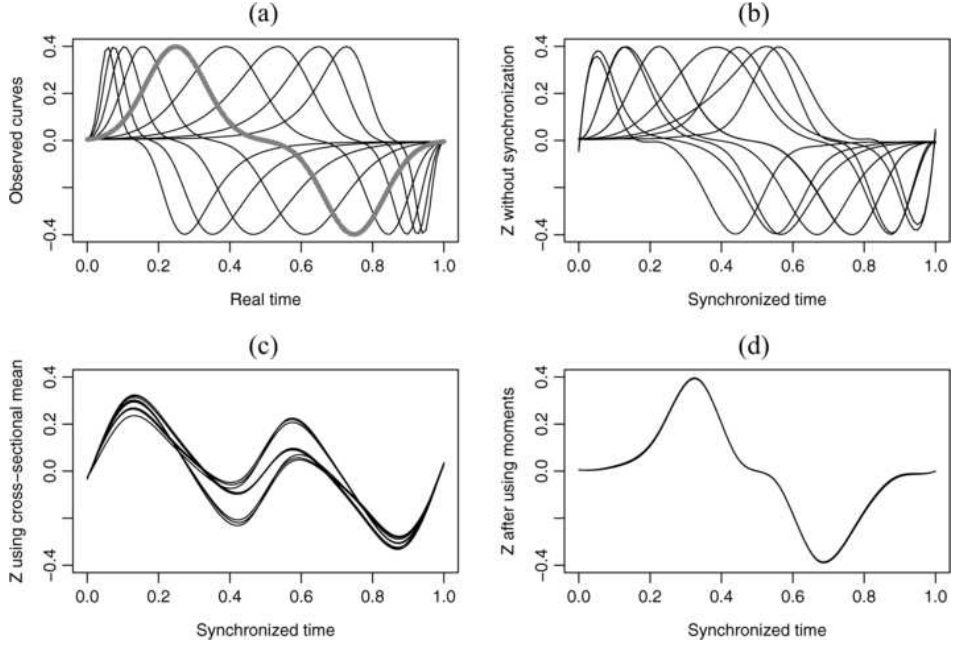

FIG. 2. (a) *A simulated set of ten curves that have been "warped" in time with the grey line indicating the original shape.* (b) *The estimates for $Z_i(t)$ with $\lambda_{\text{sync}} = \lambda_{\text{mom}} = \lambda_{\text{W}} = 0$.* (c) *Estimates for $Z_i(t)$ with $\lambda_{\text{mom}} = 0$.* (d) *Estimates including all four terms.*

Figure 2 uses a simulated data set to illustrate the need for all four terms in (7). Figure 2(a) plots ten curves, each generated from the solid grey curve in the center and then "warped" by distorting the time axis by differing amounts. Figure 2(b) illustrates the corresponding ten estimates for the $Z_i$'s, representing the "synchronized" curves, obtained by minimizing (7) with $\lambda_{\text{sync}} = \lambda_{\text{mom}} = \lambda_{\text{W}} = 0$. The fit is very good, with the estimated standard deviation of the $\varepsilon_{ij}$'s only 0.006, but this approach has clearly done a poor job of synchronizing the data. Alternatively, Figure 2(c) shows the results using $\lambda_{\text{sync}} = 10$, a small value for $\lambda_{\text{W}}$ and $\lambda_{\text{mom}} = 0$. A high level of synchronization has resulted from the use of $P(\boldsymbol{\theta}_i)$, but the curves bear little relationship to the original ones. In addition, the $Z_i$'s have been shrunk toward zero, resulting in a ten fold increase in the standard deviation of the estimates. As $\lambda_{\text{sync}}$ is reduced and $\lambda_{\text{W}}$ increased, the fit shifts toward that shown in Figure 2(b), but at no stage do we get strong synchronization, the correct shape and a good fit to the data. Finally, Figure 2(d) provides a plot of the ten estimated $Z_i$'s after setting $\lambda_{\text{mom}} > 0$ and using two moments corresponding to $I^{\max}$ and $I^{\min}$. Notice that the addition of $P(\mu_{Z_i})$ has enabled us to not only synchronize the data but to also reproduce the original shape of the curves. In addition, the estimated standard deviation is almost identical to that from the fit illustrated in Figure 2(b), indicating that the synchronization has not been at the expense of an accurate fit to the data.

There are two reasons for the inadequate fit in Figure 2(c). First, because of the significant distortion of the observed $Y_i$'s, the cross-sectional mean, which is used to compute $\boldsymbol{\mu_\theta}$, is a poor estimate for the true shape, so the curves have been synchronized toward the wrong "target." This is the same problem that one would encounter when using the continuous monotone registration approach on this data. Second, the act of shrinking has resulted in a very poor fit to the original curves. Utilizing $P(\mu_{Z_i})$ has three advantages which allows us to address both these problems. First, since the moments are measures of shifts in the time axis, forcing the $Z_i$'s to have similar moments has no effect on their amplitude and, hence, does not cause the shrinkage problem observed in Figure 2(c). Second, the moments are only a summary of each curve so can often be much more accurately estimated than the entire curve. For example, the cross-sectional mean of the curves in Figure 2(a) is a poor estimate of the overall shape of the curves. However, $\mu_{\bar{Y}}^{\max}$ and $\mu_{\bar{Y}}^{\min}$ still provide good estimates for the maximum and minimum of the original curve that the data was generated from, so the problem of aligning the curves to the wrong shape can be eliminated. Finally, one can choose among a wide range of feature functions when producing the moments. Hence, one can identify specific characteristics or features in the curves and design feature functions accordingly. Since feature functions can theoretically be designed to identify, and hence synchronize toward any consistent marker events, the landmark registration approach can be seen as a special case of the moments method.

4.2. *Asymptotic theory.* Section 5 illustrates the moments method's practical performance on the Berkeley growth curve data and Section 6 provides a comprehensive comparison to other methods on several simulated data sets. However, we can also show that, under general regularity conditions, the method exhibits good large sample properties in terms of asymptotic consistency of the estimators. Let $\boldsymbol{\eta}_0$ represent the set of parameters for our model, that is, $\boldsymbol{\gamma}_1, \ldots, \boldsymbol{\gamma}_N$ and $\boldsymbol{\theta}_1, \ldots, \boldsymbol{\theta}_N$, and $\hat{\boldsymbol{\eta}}_n$ the corresponding estimates from minimizing (7). Then we first introduce four assumptions:

(A-1) $\mu_{Z_i}^{(l,k)}$ is a continuous function of $\boldsymbol{\theta}_i$ for all $l$ and $k$. Also, $\mathbf{z}(W_i(t_j))$ is a continuous function of $\boldsymbol{\gamma}_i$.
(A-2) $\mathbf{z}(W(t))^T \boldsymbol{\theta}$ is a uniformly continuous function of $t$, that is, for all $\delta_1 > 0$, there exists $\delta_2 > 0$ such that for all $t_1, t_2$, where $|t_1 - t_2| < \delta_2$, it is the case that $|\mathbf{z}(W(t_1))^T \boldsymbol{\theta} - \mathbf{z}(W(t_2))^T \boldsymbol{\theta}| < \delta_1$ for any $\boldsymbol{\theta}$ and $\boldsymbol{\gamma}$.
(A-3) We choose feature functions and corresponding moments such that the synchronization model given by (3) and (4) is identifiable when the curves are observed over a finite set of time points, **t**.

(A-4) The parameter space is bounded, that is, $\|\boldsymbol{\eta}\|^2 < M$ for some finite $M$.

We can not hope to have consistent estimators without (A-1) because that would imply that estimating $\mu_{Z_i}^{(l,k)}$ and $\mathbf{z}(W_i(t_j))$ well did not necessarily correspond to estimating the true parameters well. (A-2) places a restriction on the lack of smoothness of the fit. Some level of smoothness must always be imposed on such fits or a line that interpolated the observed values of $Y$ would minimize the criterion. (A-3) is obviously necessary because if the model is unidentifiable we could not select the correct parameters even if we had complete information. (A-4) assumes that the estimators are not allowed to diverge off to infinity. Subject to these four assumptions, we provide the following consistency result.

THEOREM 2. *Let $\lambda_{\text{sync},n}, \lambda_{\text{W},n}$ and $\lambda_{\text{mom},n}$ represent the tuning parameters as a function of $n$. Suppose that* (A-1) *through* (A-4) *hold, that $\lambda_{\text{sync},n}$ and $\lambda_{\text{W},n}$ are $o(n)$ and that $\lambda_{\text{mom},n}$ is $O(n)$. Then $\hat{\boldsymbol{\eta}}_n$ will be a consistent estimator for $\boldsymbol{\eta}_0$, that is, $\hat{\boldsymbol{\eta}}_n \to \boldsymbol{\eta}_0$ a.s.*

4.3. *Selection of tuning parameters.* A key component of our synchronization approach is the choice of the tuning parameters $\lambda_{\text{sync}}$, $\lambda_{\text{W}}$ and $\lambda_{\text{mom}}$. The choice of these parameters is governed by a tradeoff between quality of fit, that is, how well the estimated curves fit the observed data, the level of synchronization achieved and the amount of distortion to the shape in performing the synchronization. In general, improving performance in one of these characteristics will cause a deterioration in the other two. An analogy would be choosing between small probabilities of type 1 and type 2 errors in hypothesis tests. Of course, the standard approach in that setting is to minimize the probability of a type 2 error subject to an upper bound constraint on the probability of a type 1 error. We take a similar approach here by selecting the tuning parameters to produce the best possible synchronization subject to constraints on the lack of fit and the distortion of the shape.

We measure the level of synchronization, $Sync$, using the average squared deviation of the synchronized curves from their mean curve as a percentage of the same quantity for the unsynchronized curves. Hence, a value of zero would indicate an identical shape for all synchronized curves, while one corresponds to no improvement in the synchronization. The lack of fit, $\sigma$, is quantified using the average standard deviation between the observed curves, $Y_i(t_j)$, and their "estimates," $\hat{Y}_i(t_j)$. Finally, the distortion to the shape of the curves is measured using $P(W)$. We then select the tuning parameters so as to minimize $Sync$ subject to $\sigma$ and $P(W)$ being less than certain upper bounds. Performing this optimization over three parameters is a potentially difficult computational task. Fortunately, the fit turns out to be fairly stable for wide ranges of possible values for $\lambda_{\mathrm{W}}$ and $\lambda_{\mathrm{mom}}$, while $\lambda_{\mathrm{sync}}$ has a considerably stronger influence. In the case of $\lambda_{\mathrm{mom}}$ it makes intuitive sense that its exact value is not important because the moments are acting to produce an identifiable result, so any reasonable weight will make the model identifiable and, hence, produce a good fit. Hence, it is feasible to implement a grid search over the three parameters where the grid for $\lambda_{\mathrm{W}}$ and $\lambda_{\mathrm{mom}}$ is very coarse, while the grid for $\lambda_{\mathrm{sync}}$ needs to be considerably finer. For the growth curve data, illustrated in the following section, we use values of $10^3$, $10^4$, $10^5$ and $10^6$ for $\lambda_{\mathrm{mom}}$ and values of $10^{-1}$, $10^0$ and $10^1$ for $\lambda_{\mathrm{W}}$. We have found these grids to work well for the problems we have examined. This is consistent with Ramsay and Li (1998) who also found that a small grid of tuning parameters worked over a wide range of applications. In theory cross-validation could be used to select the dimensions of the basis functions **z** and **w**. However, in practice we have found that, given the flexibility provided by the three tuning parameters, any dimension that provides a reasonably flexible basis will suffice.

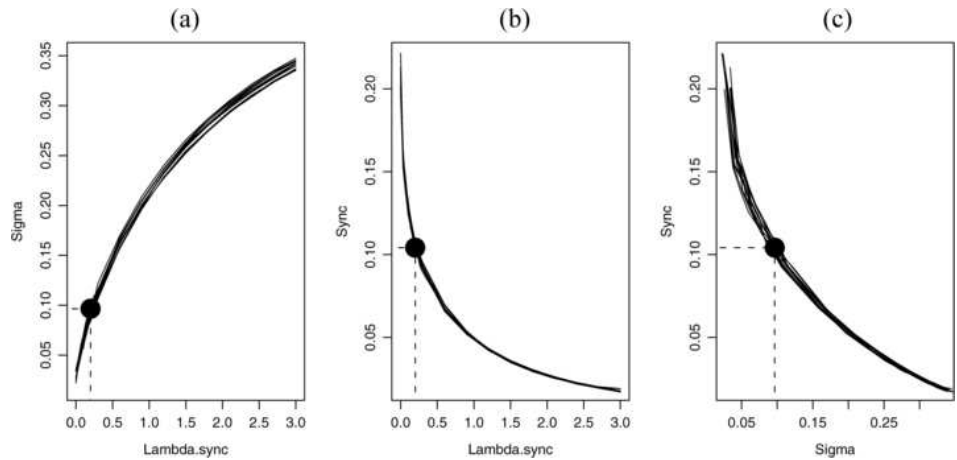

Fig. 3. *Plots of* (a) $\sigma$ *versus* $\lambda_{\mathrm{sync}}$, (b) *Sync versus* $\lambda_{\mathrm{sync}}$ *and* (c) *Sync versus* $\sigma$ *for four different values of* $\lambda_{\mathrm{mom}}$ *and three different values for* $\lambda_{\mathrm{W}}$.

**5. An application to the Berkeley growth curve data.** In this section we demonstrate the moments based method on the Berkeley growth curve data, discussed in Section 1, utilizing the nonlinear warping functions, $W_i$, given by (6). The data were obtained by fitting a smoothing spline to the second differences of the observed heights for each of ten boys. The smoothing was performed to aid visualizing the resulting curves. We have also performed registration on the raw data with similar results. The first step in implementing our approach involves the choice of the feature functions. This data exhibits clear global maximums and minimums so we elected to utilize $I_g^{\max}$ and $I_g^{\min}$ with $r = 100$. For both feature functions we concentrated on the first moment, but one could also have used additional higher-order moments. Next we selected the tuning parameters using the approach from Section 4.3. Figure 3 provides an illustration of this method. Each plot contains 12 separate lines corresponding to four different values for $\lambda_{\mathrm{mom}}$ ($10^3$, $10^4$, $10^5$, $10^6$) and three different values for $\lambda_{\mathrm{W}}$ ($10^{-1}$, $10^0$, $10^1$). The 12 lines are almost indistinguishable from each other, emphasizing the insensitivity of the result to the exact choice of $\lambda_{\mathrm{mom}}$ and $\lambda_{\mathrm{W}}$. Figure 3(a) plots $\sigma$ as a function of $\lambda_{\mathrm{sync}}$. Similarly, Figure 3(b) plots *Sync* as a function of $\lambda_{\mathrm{sync}}$. Finally, Figure 3(c) plots *Sync* as a function of $\sigma$. All three plots show a smooth tradeoff between $\sigma$ and *Sync* with little effect from the other two tuning parameters. We opted to use tuning parameters that produced the optimal synchronization subject to $\sigma$ being no larger than 0.1 and $P(W)$ no greater than 0.5. These cutoffs were chosen because they seemed to produce a high level of synchronization with a relatively low increase in $\sigma$. The dots on Figure 3 correspond to this fit ($\lambda_{\mathrm{sync}} = 0.2$, $\lambda_{\mathrm{W}} = 10$, $\lambda_{\mathrm{mom}} = 10^5$). We can see that attempting to further synchronize the curves past this point will result in a large increase in $\sigma$.

Figure 1(b) in Section 1 provides a plot of the synchronized curves, $Z_i$, from the resulting fit. Notice that the synchronized mean curve not only appears to estimate the correct height for the peaks and troughs but also shifts the peak to a later age from that of the cross-sectional mean. To help judge the accuracy of our procedure, Figure 4 provides a comparison to other potential methods. Here we have plotted the estimated mean acceleration curve using five different approaches. In particular, we applied our moments method using the above tuning parameters, the moments method with $\lambda_{\mathrm{mom}} = 0$, landmark registration (aligning on the peak and the trough of each curve), the continuous monotone registration method and the cross-sectional mean from the unaligned curves. The cross-sectional mean is well known to be inadequate for this data set [Gasser et al. (1984)]. However, the landmark method provides a natural gold standard for this problem because it is known to work extremely well in situations such as this one where each curve exhibits a very similar structure [Kneip and Gasser (1992)]. All four methods give considerable improvements over the cross-sectional mean, but the moments method with $\lambda_{\mathrm{mom}} = 10^5$ gives the most similar fit to the landmark approach. The continuous monotone registration method gives the worst performance of the four because it does not take advantage of the specific shape information in the data. Finally, the moments method with $\lambda_{\mathrm{mom}} = 0$ gives somewhat intermediate performance. While it does a good job correctly estimating the trough, it fails to identify the correct location of the peak. Again, this is because it fails to make full use of the structure that is present. This illustrates that, while the results are relatively insensitive to the choice of $\lambda_{\mathrm{mom}}$, this term is still a vital part of the fit.

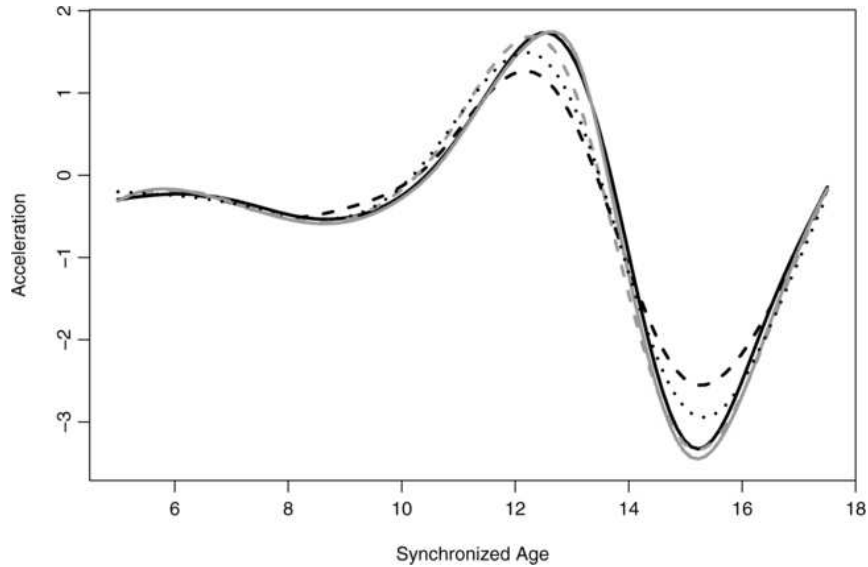

Fig. 4. *Plots of mean curves on the Berkeley growth curve data using cross-sectional mean (dashed black), continuous monotone registration (dotted), moments method with $\lambda_{\mathrm{mom}} = 0$ (dashed grey) and $\lambda_{\mathrm{mom}} = 10^5$ (solid grey), and landmark registration (solid black).*

**6. Simulation study.** In this section we compare the performance of our moments based synchronization approach with the continuous monotone registration and landmark methods over four sets of simulations. For each simulation 100 data sets, each consisting of ten curves sampled at 100 equally spaced time points, were generated from a given distribution. Six different synchronization methods were then applied to each data set corresponding to the moments, continuous monotone registration and landmark procedures using both linear and standardized, (6), warping functions. For the moments method, we used $K = 1$ moment for each feature function. For each set of simulations, the $\lambda$ parameters were chosen by selecting the values that provided maximum alignment on a preliminary data set subject to constraints on $\sigma$ and $P(W)$ as discussed in Section 4.3. The simulation results are summarized in Table 1. Two numbers are provided for each simulation-method pair corresponding to *Sync* and $\sigma$ as defined in Section 4.3. For the moments method, $\sigma$ was produced using $Z_i(W_i(t))$, while for the other two methods it was computed using a smoothing of the curves performed via a smoothing spline prior to synchronization.

Simulation one consisted of curves generated from a standard Gaussian density which were then stretched and shifted in the $X$ or time axis. Figure 5(a) illustrates a typical set of curves. We used the peak of each curve as the marker event for the landmark methods and $I^{\max}(t)$ ($r = 100$) for the moments methods. For this simulation, the continuous monotone registration and moments methods both worked very well. In particular, the continuous monotone registration method produced good results because the cross-sectional mean of the observed curves, used to produce the target function, still had an approximate bell shape. There was little difference between the linear and nonlinear warping functions because the true warping was in fact linear. The landmark method, while still providing a considerable level of synchronization, performed relatively less well because, with only one marker, it could not adequately correct for differences in the spread of the curves.

TABLE 1
*Results from four simulations on six different alignment methods. Sync is measured as a percentage so* 100 *corresponds to no improvement in synchronization. The standard errors on the Sync were between* 0.15 *and* 0.45 *for those results marked with an* $^*$ *and were less than* 0.15 *for all others*

| $W(t)$ | Method | Simulation | | | | | | | |
|---|---|---|---|---|---|---|---|---|---|
| | | One | | Two | | Three | | Four | |
| | | Sync | $\sigma$ | Sync | $\sigma$ | Sync | $\sigma$ | Sync | $\sigma$ |
| Linear | Cont. Mono. Reg. | 0.02 | 0.0005 | 76.50 | 0.0003 | 78.63$^*$ | 0.004 | 75.37$^*$ | 0.009 |
| | Landmark | 11.86$^*$ | 0.0005 | 1.15 | 0.0003 | 9.39 | 0.004 | 15.64 | 0.009 |
| | Moments | 0.07 | 0.0013 | 0.50 | 0.0020 | 8.33$^*$ | 0.028 | 13.71$^*$ | 0.039 |
| Nonlinear | Cont. Mono. Reg. | 0.06 | 0.0005 | 39.47 | 0.0003 | 21.55$^*$ | 0.004 | 21.18 | 0.009 |
| | Landmark | 12.32$^*$ | 0.0005 | 6.31 | 0.0003 | 2.86 | 0.004 | 1.42 | 0.009 |
| | Moments | <0.01 | 0.0013 | 0.59 | 0.0008 | 0.76 | 0.009 | 1.20 | 0.014 |

Simulation two had a similar set up to the previous simulation except that half the curves were centered close to 0.7, while the others were centered close to 0.3 [see Figure 5(b)]. As a result, the cross-sectional mean was bimodal, which significantly adversely affected the continuous monotone registration method. The landmark method performed relatively better on this data because shifts in the curve, which it could correct for, formed a larger portion of the lack of synchronization. The moments method was only marginally affected by the bimodal shape of the data.

For simulation three we generated curves using the distribution illustrated in Figure 2(a). These curves were produced using a nonlinear warping function and presented a more challenging problem. We utilized both the maximum and minimum points as markers for the landmark methods and $I^{\max}(t)$ and $I^{\min}(t)$ ($r = 100$) for the moments methods. Again, the continuous monotone registration method performed poorly because the cross-sectional mean did not adequately reflect the shape of the curves. The landmark and moments methods both gave good results. For all three procedures the nonlinear warping functions worked considerably better than their linear counterparts. Finally, the fourth simulation tested out the effect of noise in the observed curves by adding Gaussian errors with standard deviation of 0.01 to the data from simulation three [see Figure 5(c)]. We also added a linear drift in the curves to ensure that the moments method still performed well when the curves started and ended at differing values on the $Y$-axis. In general, these changes caused a moderate deterioration in the linear versions of the landmark and moments procedures, presumably because the drift in the curves made it harder for a linear warping function to accurately realign the curves. However, the nonlinear versions gave fairly similar performance to those of simulation three. Note that some improvement in the moments method results may have been possible if we had smoothed the curves before applying our approach. However, given the small deterioration from simulation three, it is doubtful that any significant gains would have been achieved.

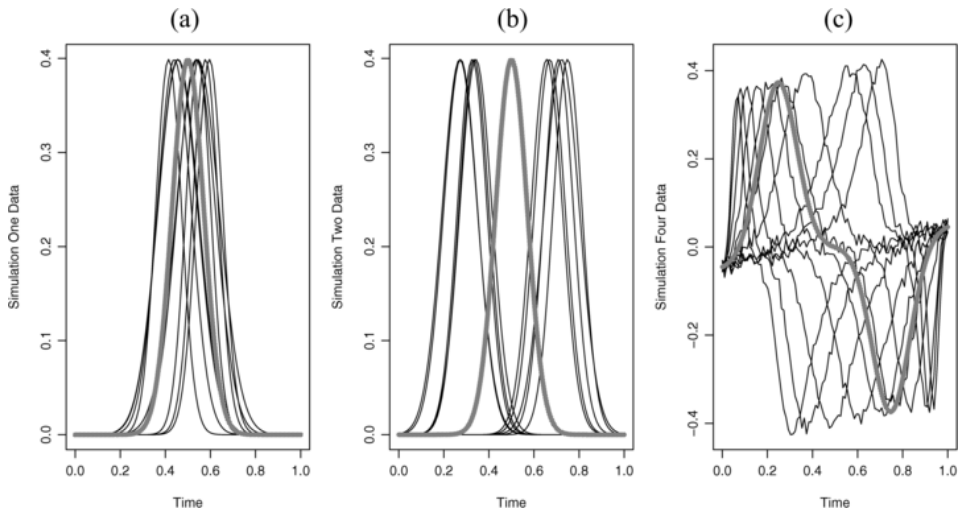

Fig. 5. (a) *A simulated set of ten curves that have been "warped" in time. This is one of* 100 *data sets from simulation one.* (b) *One of the* 100 *data sets from simulation two.* (c) *One of the* 100 *data sets from simulation four. For each plot, the thicker grey line indicates the original shape.*

These simulation results may be somewhat unfair to the landmark method because it is difficult to implement this approach in a truly automatic fashion. For example, by manually identifying additional landmarks in the simulated curves, one may have been able to produce fits closer to that from the moments approach. However, our attempt here is not necessarily to show that our approach will outperform landmark registration where multiple marker events can be manually identified, since landmark registration is considered the benchmark in this case. Rather, we want to show that the moments method can give comparable results, without the need for manual intervention, when marker events are present, but can also provide accurate results even in the absence of such markers.

Notice that because of the way that the moments method works its $\sigma$ was somewhat higher on all four simulations than for the continuous monotone registration or landmark methods. This is one of the tradeoffs for a higher level of synchronization. However, the increase is relatively small, particularly for the nonlinear warping functions, so the tradeoff clearly seems worthwhile. Simulations two and three illustrate the advantage of combining landmark and continuous monotone registration criteria together. By first synchronizing based on landmarks, such as turning points, we can achieve a partial synchronization and then estimate $\boldsymbol{\mu_\theta}$ well enough to produce a very accurate final alignment. In such situations we have found that the best results are obtained by using a relatively higher value for $\lambda_{\text{mom}}$ in the first few iterations and then reducing $\lambda_{\text{mom}}$ while increasing $\lambda_{\text{sync}}$ in the remaining iterations. This is the approach we took for these simulations.

**7. Discussion.** In this article we have developed a general moments based approach to the problem of synchronization of functional or curve data. The generally accepted benchmark for such problems is landmark registration which aligns curves by identifying marker events. This approach can be very effective but has two, potentially significant, disadvantages. First, it assumes all curves have consistent marker events and, second, even if the marker events exist, one often must manually identify them, which is not feasible for large data sets. Alternatively, the continuous monotone registration method works well when an adequate target function, $T(t)$, can be identified but fails when the data is poorly enough aligned that $T(t)$ does not match the shape of the curves. The moments based approach builds on the strengths of both methods and reduces or eliminates their deficiencies. As with the landmark approach, for those curves with marker events, feature functions, such as $I^{\max}(t)$ or $I^{\min}(t)$, can be implemented to synchronize based on these events. However, for curves, or data sets, that do not exhibit such markers, more global feature functions, such as $I^{(m)}(t)$, can be utilized. In this sense our method is an extension of landmark registration. When comparing to the continuous monotone registration approach, notice that $\bar{Z}(t) = \mathbf{z}(t)^T \boldsymbol{\mu_\theta}$ can be considered to be the analog of $T(t)$ in that we, at least partially, synchronize the curves toward it. However, even in situations where the cross-sectional mean provides a poor estimate for $T(t)$ and, hence, the continuous monotone registration method fails, the moments will often induce an accurate enough initial synchronization that $\bar{Z}(t)$ will represent the correct shape. Hence, as the method iterates through the fitting algorithm, the synchronization becomes better as opposed to the continuous monotone registration fit where no improvement may be possible. The data from simulations two and three provide a good illustration of this effect. Hence, our approach can also be considered as an extension of continuous monotone registration.

This method could be generalized in several directions. Although, in this article, we have only discussed one-dimensional curves, the moments approach could potentially be extended to multidimensional data. The definition of the feature function, $I_g(t)$, could easily be expanded to such data and hence the moments also. Equating the lower-order moments could then be achieved in a similar fashion to the one-dimensional case. The most significant challenge would seem to be dealing with higher-order moments on high-dimensional data where the number of cross product terms could become unmanageable. Another possible extension is to attempt to model the covariance of the $\boldsymbol{\theta}_i$'s, $\text{Var}(\boldsymbol{\theta}_i) = \Theta$. For example, $P(\boldsymbol{\theta}_i)$ could be altered to include $\Theta$ using, $P^*(\boldsymbol{\theta}_i) = (\boldsymbol{\theta}_i - \boldsymbol{\mu_\theta})^T \Theta^{-1} (\boldsymbol{\theta}_i - \boldsymbol{\mu_\theta})$. There are several possible ways to model $\Theta$. The first, which we have effectively used in $P(\boldsymbol{\theta}_i)$, is to take $\Theta$ equal to a multiple of the identity matrix. One could also estimate $\Theta$ at each iteration via the sample covariance, $\hat{\Theta} = \frac{1}{N}\sum_i (\boldsymbol{\theta}_i - \boldsymbol{\mu_\theta})(\boldsymbol{\theta}_i - \boldsymbol{\mu_\theta})^T$. However, such an unconstrained estimate may be impractical if the dimension of the $\boldsymbol{\theta}_i$'s is large. One solution would be to constrain the rank of $\Theta$ and, hence, significantly reduce the number of parameters to estimate [James, Hastie and Sugar (2000)]. Another appealing alternative would be to design $\Theta$ such that $P^*(\boldsymbol{\theta}_i)$ placed no penalty on values of $\boldsymbol{\theta}_i$ corresponding to constant vertical shifts of $\mathbf{z}(t)^T \boldsymbol{\theta}_i$. This would mean that two curves that differed only by a constant vertical shift would be considered to be perfectly synchronized and would likely significantly reduce the undesirable shrinkage toward the mean that, for example, is evident in Figure 2(c).

## APPENDIX

**A.1. Proof of Theorem 1.** First note that (2) implies that

$$I_{g((s-a)/b)}(t) = \frac{I_g((t-a)/b)}{\int I_g((t-a)/b)\,dt} = \frac{1}{b} I_g\left(\frac{t-a}{b}\right).$$

Hence,

$$\mu_{h((s-a)/b)}^{(1)} = \int t I_{h((s-a)/b)}(t)\,dt = \int t \frac{1}{b} I_h\left(\frac{t-a}{b}\right) dt$$

$$= \int (sb+a) I_h(s)\,ds = b\int I_h(s)\,ds + a\int I_h(s)\,ds = b\mu_h^{(1)} + a,$$

where $s = \frac{t-a}{b}$. Similarly,

$$\mu_{h((s-a)/b)}^{(k)} = \int (t - b\mu_h^{(1)} - a)^k I_{h((s-a)/b)}(t)\,dt$$

$$= \int \frac{1}{b}(t - b\mu_h^{(1)} - a)^k I_h\left(\frac{t-a}{b}\right) dt$$

$$= \int (sb - b\mu_h^{(1)})^k I_h(s)\, ds = b^k \mu_h^{(k)}.$$

**A.2. Proof of Corollary 1.** First note that if $I_g^\phi(t) \propto \phi(g(t))$, then $I_{g((s-a)/b)}^\phi(t) \propto \phi(g(\frac{t-a}{b})) \propto I_g^\phi(\frac{t-a}{b})$. Next note that

(8) $$\frac{d^m g((t-a)/b)}{dt^m} = \frac{1}{b^m} g^{(m)}\left(\frac{t-a}{b}\right),$$

so $I_{g((s-a)/b)}^{(m)}(t) \propto |g^{(m)}(\frac{t-a}{b})| \propto I_g^{(m)}(\frac{t-a}{b})$. To show the result for $I_g^{\max}$ note that

$$I_{g((s-a)/b)}^{\max}(t) \propto \left(g\left(\frac{t-a}{b}\right) - \min\left\{g\left(\frac{t-a}{b}\right)\right\}\right)^\delta$$

$$= \left(g\left(\frac{t-a}{b}\right) - \min\{g(t)\}\right)^\delta \propto I_g^{\max}\left(\frac{t-a}{b}\right)$$

and similarly for $I_g^{\min}$. Finally, by (8),

$$\frac{dg((t-a)/b)/dt}{\sqrt{d^2 g((t-a)/b)/dt^2}} = \frac{g^{(1)}((t-a)/b)/b}{\sqrt{g^{(2)}((t-a)/b)/b^2}} = \frac{g^{(1)}((t-a)/b)}{\sqrt{g^{(2)}((t-a)/b)}},$$

so

$$I_{g((s-a)/b)}^{\text{local}}(t) \propto \exp\left(-\delta \frac{dg((t-a)/b)/dt}{\sqrt{(d^2 g((t-a)/b))/dt^2}}\right)$$

$$\propto \exp\left(-\delta \frac{g^{(1)}((t-a)/b)}{\sqrt{g^{(2)}((t-a)/b)}}\right) \propto I_g^{\text{local}}\left(\frac{t-a}{b}\right).$$

**A.3. Proof of Theorem 2.** First we state and prove a lemma.

LEMMA A.1. *Suppose*

(9) $$\sup_t |\mathbf{z}(\hat{W}_n(t))^T \hat{\boldsymbol{\theta}}_n - \mathbf{z}(W_0(t))^T \boldsymbol{\theta}_0| \to 0 \quad a.s.$$

*and*

(10) $$\mu_{\hat{Z}_n}^{(l,k)} \to \mu_{\bar{Y}}^{(l,k)} \quad a.s. \text{ for } l = 1, \ldots, L \text{ and } k = 1, \ldots, K_l,$$

*where $\hat{Z}_n(t) = \mathbf{z}(t)^T \hat{\boldsymbol{\theta}}_n$ and $\hat{W}_n$ and $W_0$ respectively represent the warping functions evaluated at $\hat{\boldsymbol{\gamma}}_n$ and $\boldsymbol{\gamma}_0$. Then, provided* (A-1), (A-3) *and* (A-4) *hold, $\hat{\boldsymbol{\theta}}_n \to \boldsymbol{\theta}_0$ a.s. and $\hat{\boldsymbol{\gamma}}_n \to \boldsymbol{\gamma}_0$ a.s.*

A.3.1. *Proof of Lemma* A.1. Note we treat each curve individually, so we drop the subscript $i$ and let $\boldsymbol{\eta}_0 = \binom{\boldsymbol{\gamma}_0}{\boldsymbol{\theta}_0}$ and $\hat{\boldsymbol{\eta}}_n = \binom{\hat{\boldsymbol{\gamma}}_n}{\hat{\boldsymbol{\theta}}_n}$. To reduce notation, let

$$f(\boldsymbol{\eta}, t) = \mathbf{z}(W(t))^T \boldsymbol{\theta}.$$

First, note that (9) and (10) imply that there exists $\Omega^*$ with $P(\Omega^*) = 1$ s.t. $\forall \omega^* \in \Omega^*$,

$$f(\hat{\boldsymbol{\eta}}_n(\omega^*), t) \to f(\boldsymbol{\eta}_0, t) \qquad \forall t \tag{11}$$

and

$$\mu^{(l,k)}_{\hat{Z}_n}(\omega^*) \to \mu^{(l,k)}_{\bar{Y}} \qquad \text{for } l = 1, \ldots, L \text{ and } k = 1, \ldots, K_l. \tag{12}$$

Now, suppose that $\hat{\boldsymbol{\eta}}_n$ does not converge a.s. to $\boldsymbol{\eta}_0$. This implies there exists $\Omega$ with $P(\Omega) > 0$ s.t. $\forall \omega \in \Omega$, $\hat{\boldsymbol{\eta}}_n(\omega)$ does not converge to $\boldsymbol{\eta}_0$. Since the intersection of $\Omega^*$ and $\Omega$ must be nonempty, we take a particular $\omega \in \Omega^* \cap \Omega$. Then there exists an infinite subsequence $n'(\omega)$ and $\delta(\omega) > 0$ such that

$$\|\hat{\boldsymbol{\eta}}_{n'(\omega)}(\omega) - \boldsymbol{\eta}_0\| > \delta(\omega) \tag{13}$$

for all $n'(\omega)$. But recall that any bounded sequence must have a convergent subsequence. Hence, by boundedness of $\boldsymbol{\gamma}$ and $\boldsymbol{\theta}$, (A-4), there must be a subsequence, $n''(\omega)$, of $n'(\omega)$, and a $\boldsymbol{\eta}^*(\omega)$, such that

$$\hat{\boldsymbol{\eta}}_{n''(\omega)}(\omega) \to \boldsymbol{\eta}^*(\omega). \tag{14}$$

Let $W^*$ represent the warping function evaluated at $\boldsymbol{\gamma}^*$. Then, since $\mathbf{z}(\hat{W}_n)$ is a continuous function of $\hat{\boldsymbol{\gamma}}_n$ and $\mu^{(l,k)}_{\hat{Z}_n}$ is a continuous function of $\hat{\boldsymbol{\theta}}_n$ [by (A-1)], $f$ is continuous and, hence, (14) implies that

$$f(\hat{\boldsymbol{\eta}}_{n''(\omega)}(\omega), t) \to f(\hat{\boldsymbol{\eta}}^*(\omega), t) \qquad \forall t \tag{15}$$

and

$$\mu^{(l,k)}_{\hat{Z}_n}(\omega) \to \mu^{(l,k)}_{Z^*} \qquad \text{for } l = 1, \ldots, L \text{ and } k = 1, \ldots, K_l. \tag{16}$$

Now, (11) and (15) imply that $f(\boldsymbol{\eta}_0, t) = f(\hat{\boldsymbol{\eta}}^*(\omega), t)$ for all $t$, while (12) and (16) imply that $\mu^{(l,k)}_{\bar{Y}} = \mu^{(l,k)}_{Z^*}$ for $l = 1, \ldots, L$ and $k = 1, \ldots, K_l$. By moments identifiability of the model, (A-3), this implies $\boldsymbol{\eta}^*(\omega) = \boldsymbol{\eta}_0$. But by (13) and (14), $\|\boldsymbol{\eta}^*(\omega) - \boldsymbol{\eta}_0\| > 0$, which is a contradiction. Hence, $\hat{\boldsymbol{\eta}}_n \to \boldsymbol{\eta}_0$ a.s.

A.3.2. *Proof of the theorem.* Let $\boldsymbol{\eta}$ represent the set of parameters for our model, that is, $\boldsymbol{\gamma}_1, \ldots, \boldsymbol{\gamma}_N, \boldsymbol{\theta}_1, \ldots, \boldsymbol{\theta}_N$, and $\boldsymbol{\theta}_\mu$. Each curve is evaluated at $n$ time points, $t_1, \ldots, t_n$. Let

$$a_n(\boldsymbol{\eta}) = \frac{1}{n}\sum_{i=1}^{N}\sum_{j=1}^{n}(Y_{ij} - \mathbf{z}(W_{ij})^T\boldsymbol{\theta}_i)^2,$$

$$b(\boldsymbol{\eta}) = \lambda_{\text{mom}}\frac{1}{n}\sum_{i=1}^{N}\sum_{l=1}^{L}\sum_{k=1}^{K_l}(\mu_{Z_i}^{(l,k)} - \mu_{\bar{Y}}^{(l,k)})^2,$$

$c(\boldsymbol{\eta}) = \sum_{i=1}^{N}\|\boldsymbol{\theta}_i - \boldsymbol{\mu}_{\boldsymbol{\theta}}\|^2$ and $d(\boldsymbol{\eta}) = \sum_{i=1}^{N}P(W_i)$, where $W_{ij} = W_{\gamma_i}(t_j)$. So $Q_n(\boldsymbol{\eta}) = a_n(\boldsymbol{\eta}) + b(\boldsymbol{\eta}) + \frac{\lambda_{\text{sync},n}}{n}c(\boldsymbol{\eta}) + \frac{\lambda_{\text{W},n}}{n}d(\boldsymbol{\eta})$ represents (7) using $n$ time points. Let $\boldsymbol{\eta}_0$ represent the true parameters and $\hat{\boldsymbol{\eta}}_n$ the estimators resulting from minimizing $Q_n$. Then $Q_n(\boldsymbol{\eta}_0) = a_n(\boldsymbol{\eta}_0) + \frac{\lambda_{\text{sync},n}}{n}c(\boldsymbol{\eta}_0) + \frac{\lambda_{\text{W},n}}{n}d(\boldsymbol{\eta}_0)$, where $c(\boldsymbol{\eta}_0)$ and $d(\boldsymbol{\eta}_0)$ are both finite. Note $c(\boldsymbol{\eta}_0)$ is finite since $\boldsymbol{\theta}$ is bounded and $d(\boldsymbol{\eta}_0)$ is finite because, by (5), $0 < W'_{0_i}(t) < \infty$ for $t \in [0, T]$, provided $f_{0_i}(t)$ is bounded and this is the case because $\boldsymbol{\gamma}_{0_i}$ is bounded. Also, $b(\boldsymbol{\eta}_0) = 0$ because $\mu_{Z_{0_1}}^{(l,k)} = \mu_{Z_{0_2}}^{(l,k)} = \cdots = \mu_{Z_{0_N}}^{(l,k)} = \mu_{\bar{Y}}^{(l,k)}$ for all $l$ and $k$ where $Z_{0_i} = \mathbf{z}^T\boldsymbol{\theta}_{0_i}$. Clearly, $Q_n(\hat{\boldsymbol{\eta}}_n) \leq Q_n(\boldsymbol{\eta}_0)$ because $\hat{\boldsymbol{\eta}}_n$ is optimized over all $\boldsymbol{\eta}$. Also, $Q_n(\hat{\boldsymbol{\eta}}_n) \geq a_n(\hat{\boldsymbol{\eta}}_n) + b(\hat{\boldsymbol{\eta}}_n)$ because $c$ and $d$ are positive. Hence,

$$(17) \qquad a_n(\hat{\boldsymbol{\eta}}_n) + b(\hat{\boldsymbol{\eta}}_n) \leq a_n(\boldsymbol{\eta}_0) + \frac{\lambda_{\text{sync},n}}{n}c(\boldsymbol{\eta}_0) + \frac{\lambda_{\text{W},n}}{n}d(\boldsymbol{\eta}_0).$$

Let $\phi_{n_{ij}} = (\mathbf{z}(W_{0_{ij}})^T\boldsymbol{\theta}_{0_i} - \mathbf{z}(\hat{W}_{n_{ij}})^T\hat{\boldsymbol{\theta}}_{n_i})$ and $\varepsilon_{ij} = (Y_{ij} - \mathbf{z}(W_{0_{ij}})^T\boldsymbol{\theta}_{0_i})$, where $W_{0_{ij}} = W(t_j)$ evaluated using the true $\gamma_i$ and $\hat{W}_{n_{ij}} = W(t_j)$ using $\gamma_i$ from $\hat{\boldsymbol{\eta}}_n$. Then

$$a_n(\hat{\boldsymbol{\eta}}_n) = \frac{1}{n}\sum_{i=1}^{N}\sum_{j=1}^{n}(Y_{ij} - \mathbf{z}(\hat{W}_{n_{ij}})^T\hat{\boldsymbol{\theta}}_{n_i})^2$$

$$= \frac{1}{n}\sum_{i=1}^{N}\sum_{j=1}^{n}(Y_{ij} - \mathbf{z}(W_{0_{ij}})^T\boldsymbol{\theta}_{0_i} + \mathbf{z}(W_{0_{ij}})^T\boldsymbol{\theta}_{0_i} - \mathbf{z}(\hat{W}_{n_{ij}})^T\hat{\boldsymbol{\theta}}_{n_i})^2$$

$$(18) \qquad = \frac{1}{n}\sum_{i=1}^{N}\sum_{j=1}^{n}(\varepsilon_{ij} + \phi_{n_{ij}})^2$$

$$= \frac{1}{n}\sum_{i=1}^{N}\sum_{j=1}^{n}\varepsilon_{ij}^2 + \frac{1}{n}\sum_{i=1}^{N}\sum_{j=1}^{n}\phi_{n_{ij}}^2 + 2\frac{1}{n}\sum_{i=1}^{N}\sum_{j=1}^{n}\varepsilon_{ij}\phi_{n_{ij}}$$

$$= a_n(\boldsymbol{\eta}_0) + \frac{1}{n}\sum_{i=1}^{N}\sum_{j=1}^{n}\phi_{n_{ij}}^2 + 2\frac{1}{n}\sum_{i=1}^{N}\sum_{j=1}^{n}\varepsilon_{ij}\phi_{n_{ij}}.$$

Therefore, by (17) and (18),

$$(19) \quad \frac{1}{n}\sum_{i=1}^{N}\sum_{j=1}^{n}\phi_{n_{ij}}^2 + 2\frac{1}{n}\sum_{i=1}^{N}\sum_{j=1}^{n}\varepsilon_{ij}\phi_{n_{ij}} + b(\hat{\boldsymbol{\eta}}_n) \leq \frac{\lambda_{\text{sync},n}}{n}c(\boldsymbol{\eta}_0) + \frac{\lambda_{\text{W},n}}{n}d(\boldsymbol{\eta}_0).$$

But notice that the $\varepsilon_{ij}$'s are i.i.d. mean zero random variables. Also, $\phi_{n_{ij}}$ is a difference of two bounded uniformly continuous functions, so is also bounded and uniformly continuous. [Note $\mathbf{z}(W)^T\boldsymbol{\theta}$ is uniformly continuous by (A-2) and is bounded because it is a continuous function of bounded parameters, $\boldsymbol{\gamma}$ and $\boldsymbol{\theta}$, by (A-1) and (A-4)]. Hence, by a standard application of the SLLN, $\frac{1}{n}\sum_{j=1}^{n}\varepsilon_{ij}\phi_{n_{ij}} \to 0$ a.s. as $n \to \infty$. [See Theorem 1.13(ii) in Shao (2003) for a proof of this result.] In addition, $\lambda_{\text{sync},n}$ and $\lambda_{\text{W},n}$ are $o(n)$, so the right-hand side of (19) also converges to 0. Therefore, it must be the case that

$$(20) \quad \frac{1}{n}\sum_{i=1}^{N}\sum_{j=1}^{n}\phi_{n_{ij}}^2 \to 0 \qquad \text{a.s. for all } i$$

and

$$(21) \quad b_n(\hat{\boldsymbol{\eta}}_n) \to 0 \qquad \text{a.s.}$$

Since $\lambda_{\text{mom}}$ is $O(n)$, (21) implies that (10) in Lemma A.1 must hold for each curve $i$. Finally, to show that (9) holds, we divide the time interval $[0, T]$ into $H$ equal sized regions $R_1, \ldots, R_H$. Let $n_h = n/H$ equal the number of time points in region $h$. Then, by (20), it must be the case that, for every $\omega > 0$, for large enough $n$,

$$(22) \quad \frac{1}{n_h}\sum_{j \in R_h}|\phi_{ij}| < \omega \qquad \text{a.s.}$$

But by uniform continuity, (A-2), there must be a $\delta_H > 0$ such that

$$(23) \quad |(\mathbf{z}(W_{0_i}(t))^T\boldsymbol{\theta}_{0_i} - \mathbf{z}(\hat{W}_{n_i}(t))^T\hat{\boldsymbol{\theta}}_{n_i}) - \phi_{ij}| < \delta_H$$

for any $t$ and $t_j$ in $R_h$. Combining (22) and (23), we see that

$$(24) \quad |(\mathbf{z}(W_{0_i}(t))^T\boldsymbol{\theta}_{0_i} - \mathbf{z}(\hat{W}_{n_i}(t))^T\hat{\boldsymbol{\theta}}_{n_i})| < \delta_H + \omega$$

for any $t \in R_h$ and large enough $n$. But by making $n$ large enough, this will apply simultaneously for all regions, so (24) will hold for all $t$. Now send $n \to \infty$, $H \to \infty$ and $n/H \to \infty$. Then $n_h \to \infty$ so $\omega$ can be made arbitrarily small, but also $H \to \infty$ so $\delta_H$ can also be made arbitrarily small. Hence, (9) holds for each curve $i$. Therefore, the two conditions for Lemma A.1 (9 and 10) have been proved and, therefore, by Lemma A.1, the theorem has been proved.

**Acknowledgments.** I would like to thank the Editor, Associate Editor and referees for many helpful suggestions that improved the paper.

# REFERENCES

Bar-Joseph, Z., Gerber, G. K., Gifford, D. K., Jaakkola, T. S. and Simon, I. (2003). Continuous representations of time-series gene expression data. *J. Comput. Biol.* **10** 341–356.

Ferraty, F. and Vieu, P. (2002). The functional nonparametric model and applications to spectrometric data. *Comput. Statist.* **17** 545–564. MR1952697

Ferraty, F. and Vieu, P. (2003). Curves discrimination: A nonparametric functional approach. *Comput. Statist. Data Anal.* **44** 161–173. MR2020144

Gasser, T., Müller, H.-G., Köhler, W., Molinari, L. and Prader, A. (1984). Nonparametric regression analysis of growth curves. *Ann. Statist.* **12** 210–229. MR0733509

Gervini, D. and Gasser, T. (2005). Nonparametric maximum likelihood estimation of the structural mean of a sample of curves. *Biometrika* **92** 801–820. MR2234187

James, G. M. and Hastie, T. J. (2001). Functional linear discriminant analysis for irregularly sampled curves. *J. Roy. Statist. Soc. Ser. B* **63** 533–550. MR1858401

James, G. M., Hastie, T. J. and Sugar, C. A. (2000). Principal component models for sparse functional data. *Biometrika* **87** 587–602. MR1789811

James, G. M. and Silverman, B. W. (2005). Functional adaptive model estimation. *J. Amer. Statist. Assoc.* **100** 565–576. MR2160560

James, G. M. and Sugar, C. A. (2003). Clustering for sparsely sampled functional data. *J. Amer. Statist. Assoc.* **98** 397–408. MR1995716

Kneip, A. and Gasser, T. (1992). Statistical tools to analyze data representing a sample of curves. *Ann. Statist.* **20** 1266–1305. MR1186250

Kneip, A., Li, X., MacGibbon, K. B. and Ramsay, J. O. (2000). Curve registration by local regression. *Canadian J. Statist.* **28** 19–29. MR1789833

Ramsay, J. O. and Li, X. (1998). Curve registration. *J. Roy. Statist. Soc. Ser. B* **60** 351–363. MR1616045

Ramsay, J. O. and Silverman, B. W. (2005). *Functional Data Analysis*, 2nd ed. Springer, Berlin. MR2168993

Rice, J. A. and Wu, C. O. (2001). Nonparametric mixed effects models for unequally sampled noisy curves. *Biometrics* **57** 253–259. MR1833314

Rønn, B. B. (2001). Nonparametric maximum likelihood estimation for shifted curves. *J. Roy. Statist. Soc. Ser. B Stat. Methodol.* **63** 243–259. MR1841413

Sakoe, H. and Chiba, S. (1978). Dynamic programming algorithm optimization for spoken word recognition. *IEEE Trans. Acoust. Speech Signal Processing* **26** 43–49.

Shao, J. (2003). *Mathematical Statistics*, 2nd ed. Springer, New York. MR2002723

Silverman, B. W. (1995). Incorporating parametric effects into functional principal components analysis. *J. Roy. Statist. Soc. Ser. B* **57** 673–689. MR1354074

Tuddenham, R. D. and Snyder, M. M. (1954). Physical growth of California boys and girls from birth to eighteen years. *University of California Publications in Child Development* **1** 183–364.

Zeger, S. L. and Diggle, P. J. (1994). Semiparametric models for longitudinal data with applications to CD4 cell numbers in HIV seroconverters. *Biometrics* **50** 689–699.

Department of Information and Operations Management
University of Southern California
Los Angeles, California 90089-0809
USA
E-mail: gareth@usc.edu